\documentclass{iau} 
\usepackage{graphicx}
\title[Study of AGN contribution on morphological parameters of their host galaxies] 
{Study of AGN contribution on morphological parameters of their host galaxies}
\author[Getachew-Woreta et al. 2019]   
{Tilahun Getachew-Woreta$^{1,2,3}$,
 {Mirjana Povi\'c $^{1,4}$}, {Josefa Masegosa$^4$}, {Jaime Perea$^4$}, 
 Zeleke Beyoro-Amado$^{1, 2, 5}$ \and Isabel Marquez Perez$^4$ }
\affiliation{$^1$Ethiopian Space Science and Technology Institute (ESSTI), Entoto Observatory and Research Center (EORC), Astronomy and Astrophysics Research and Development Division, P.O.Box 33679, Addis Ababa, Ethiopia\\
$^2$Addis Ababa University (AAU), P.O.Box 1176, Addis Ababa, Ethiopia \\[\affilskip]
$^3$Bule Hora University (BHU), P.O.Box 144, Bule Hora, Ethiopia \\
$^4$Instituto de Astrof\'{i}sica de Andaluc\'{i}a (IAA-CSIC), Glorieta de de la Astronom\'ia, s/n, 18008, Granada, Spain\\
$^5$Kotebe Metropolitan University (KMU), P.O.Box 31248, Addis Ababa, Ethiopia}

\pubyear{2019}
\volume{356}  
\jname{Nuclear Activity in Galaxies Across Cosmic Time}
\editors{M. Povi\'c, P. Marziani, J. Masegosa, H. Netzer,\\ S. H. Negu, \&
	S. B. Tessema, eds.}
\begin{document}

\maketitle
\begin{abstract}
We tested how the AGN contribution (5\% - 75\% of the total flux) may affect different morphological parameters commonly used in galaxy classification. We carried out all analysis at $z$\,$\sim$\,0 and at higher redshifts that correspond to the COSMOS field. Using a local training sample of $>$\,2000 visually classified galaxies, we carried out all measurements with and without the central source, and quantified
how the contribution of a bright nuclear point source could affect different morphological parameters, such as: Abraham and Concelice-Bershady indices, Gini, Asymmetry, $M20$ moment of light, and Smoothness. We found that concentration indexes are less sensitive to both redshift and brightness in comparison to the other parameters. We also found that all parameters change significantly with AGN contribution. At $z$\,$\sim$\,0, up to a 10\% of AGN contribution the morphological classification will not be significantly affect, but for $\ge$\,25\% of AGN contribution late-type spirals follow the range of parameters of elliptical galaxies and can therefore be misclassified early types.

\end{abstract}

\firstsection 
\section{Introduction}
Morphology is a key element used to study the properties of AGN host
galaxies, their connection with AGN, and their evolution (\cite[Povi\'c et al. 2012]{Povic2012}). Earlier morphological studies of local active
galaxies suggested that most of AGN reside in spiral galaxies (e.g. \cite[Adams 1977]{Adams1977}; \cite[Heckman 1978]{Heckman1978}; \cite[Filippenko 1995]{Filippenko199}; Ho 2008). Later on \cite[Kauffmann et al. (2003)]{Kauffmann2003} analysed thousands of active galaxies from the Sloan Digital Sky Survey (SDSS) at low redshifts ($z \leq 0.4$). They found that most AGN reside in massive galaxies, whose distribution of sizes, stellar surfaces, mass densities and concentration all resemble those of early-spiral SDSS galaxies. This goes in line with the results by \cite[Povi\'c et al. (2012)]{Povic2012} where in the SXDS survey (at median $z$\,$\sim$\,1.0) most of X-ray detected active galaxies were found to reside in spheroidal or bulge-dominated sources. 

The methods for morphological classification of galaxies can be separated into three categories: 1) Visual methods, used for classifying nearby and well-resolved sources. They can provide detailed information about galaxy structure, but also can be subjective and are time-consuming, especially when dealing with large
data sets (e.g., \cite[Nair $\&$ Abraham 2010]{Nair2010}, \cite[Willett et al. 2013]{Willett2013}).  2) Parametric methods, based on galaxy physical and mathematical parameters, where an analytic model for fitting the galaxy is assumed. They fail to provide again a correct description if the galaxies are not well resolved (e.g., at higher redshifts) and are time-consuming (e.g., \cite[Peng et al. 2002, 2010]{Peng2010}). 3) Non-parametric methods, that do not assume any particular analytic model, and are based on measuring different galaxy quantities that correlate with morphological types, i.e. colour, spectral properties or light distribution, galaxy shape, etc. They are less time-consuming in comparison with other methods and can provide an easy and fast separation between early- and late-type galaxies up to intermediate redshifts ($z\sim 1.5$) or higher if dealing with space-based instead of ground-based data (\cite[Povi\'c et al. 2012, 2013, 2015]{Povic2015}).

In this work, we went a step further in understanding how the presence of an AGN affects different morphological parameters commonly used in morphological classification at z\,$\sim$\,0 and at higher redshifts. We assumed a concordance cosmology with $\Omega_{\Lambda}$=0.7, $\Omega_{M} = 0.3$, and $H_{0} = 70$ km s$^{-1}$ Mpc$^{-1}$. All magnitudes given in this paper are in the $AB$ system.
\section{Data }\label{2}
\subsection{Local sample}\label{localsamples}
To study the AGN contribution on morphological parameters of active galaxies, we used  an initial sample of 8000 local galaxies in the range $0.01 \leq z \leq 0.1$ (with a mean redshift of $0.04$), observed in the SDSS (\cite[York et al. 2000]{York2000}) Data Release 4 (DR4) down to an apparent extinction-corrected magnitude of $g < 16$, and visually classified in the g-band (\cite[Nair $\&$ Abraham 2010]{Nair2010}).

The galaxies were selected randomly out of $\sim 14, 000$ sources contained in the catalogue, making sure that the selected sub-sample is a fair representation of the whole sample. The detailed description of the training sample can be found in \cite[Boyoro-Amado et al. (2019)]{Amado2019} and \cite[Povi\'c et al. (2013)]{Povic2013}.

We used the SDSS DR7 spectroscopic data (\cite[Abazajian et al. 2009]{Abazajian2009}) from the MPA-JHU emission line catalogue to select the non-AGN sample. By locating the 8000 local galaxies in the BPT diagram (\cite[Kewley et al. 2006]{Kewley2006}) we obtained that 2744 (35\%) can be classified as non-AGN, 1918 (24\%) as composite galaxies, 594 (7\%) as Seyfert2, and 2684 (34\%) as LINERs. 

After selecting the 2744 non-AGN galaxies, we further removed galaxies with
foreground stars superposed onto them and those with evident signs of interactions and mergers. Finally we ended up with a sample of 2301 non-active galaxies.
\subsection{Non-local sample (COSMOS)}\label{nonlocal}
To study morphology at higher redshift, we chose as an example the Cosmic Evolution Survey (COSMOS; \cite[Scoville et al. 2007]{Scoville2007}).  This survey is the largest project ever undertaken by the Hubble Space Telescope (HST) covering an area of 2deg$^{2}$ to a depth of $I_{ F814W} = 27.8$ mag (5$\sigma$, AB). The observations from the HST-ACS survey consist on a sample of more than $1.2 \times 10^{6}$ sources down to a magnitude limit of 26.5 in F814W filter. This field samples scale from $30-180$Mpc with $z=0.2-4$ including a million galaxies in overall volume of $10^{7}$Mpc$^{3}$. In this work we used the photometric images taken by the HST-ACS in the I$_{F814W}$ filter.
\section{Methodology and analysis}\label{descrpition}
We used the galSVM code (\cite[Huertas-Company et al. 2008]{HC2008}) for measuring the morphological parameters. It is a freely available code written as an IDL library that, combined with the similarly freely available library libSVM (\cite[{Chang $\&$ Lin} {2011}]{CL2011}) enables a morphological classification of galaxies in an automated way. This code has been especially useful when dealing with low-resolution and high redshift data (\cite[Povi\'c et al. 2015]{Povic2015}). The input information that the code needs, related to the source position, size, ellipticity parameters, and background, were obtained with SExtractor (\cite[Bertin \& Arnouts, 1996)]{Bertin1996}. The six morphological parameters used by large commonly to distinguish between early- (elliptical and lenticular) and late-type (spiral and irregular) galaxies analysed in this work are: Abraham concentration index (hereafter CABR;  \cite[Abraham et al. 1996]{Abraham1996}), Asymmetry index (A; \cite[Abraham et al. 1994]{Abraham1994}), Gini coefficient (GINI; \cite[Abraham et al. 2003]{Abraham2003}; \cite[Lotz et al. 2004]{Lotz2004}), Conselice-Bershady concentration index (CCON; \cite[Bershady et al. 2000]{Bersh2000}), $M_{20}$ moment of light (M20; \cite[Lotz et al. 2004]{Lotz2004}) and Smoothness (or clumpiness) index (\cite[Conselice et al. 2000]{Con2000}).

In this work we went through the following:
a) We measured the morphological parameters of the local sample of 2301 visually classified galaxies.
b) We obtained simulated galaxy images by adding to the original images an AGN contribution from 5-75\% of the total flux by using PSF images and a Moffat (1969) law representing the central profile. We then measured morphological parameters on all simulated images. In this way we can estimate differences on all parameters depending on the AGN contribution to the total flux.
c) We simulated the images of local sample by moving them to the rest-frame magnitude (from 21 to 25) and redshift distribution of the COSMOS field, and we measured the parameters following the procedures described in (\cite[Povi\'c et al. 2015]{Povic2015}). With this step we can test how morphological parameters change with magnitude and redshift. 
d) Finally, we repeated the previous step, but now taking into account the simulated images with different AGN contributions obtained in step b). The aim of this final step is to study the effect on morphology when having both, AGN contribution and the effect of brightness and high redshift. 
In this paper we will principally focus on the results obtained in steps a), b) and c).

\section{Results}
In this paper, we show the results obtained on the morphological parameters. Figure 1 shows the variations of the morphological parameters depending on the AGN contribution to the total flux. It can be seen that all parameters change significantly when the AGN contribution is larger than 25\% of the total flux.
Figure 2 shows the same kind of analysis when moving to higher redshifts and fainter magnitudes. The results are similar to those obtained by \cite[Povi\'c et al. 2015]{Povic2015}, although in that work no separation was done regarding the effect that the AGN can have on the morphological classification of active galaxies. The three concentration indices, CABR, CCON, and GINI show small variation when going to fainter magnitudes (typically no larger than 20\%), but ASYM and SMOOTH show to be more sensitive to both redshift and brightness.

\begin{figure}{}
\begin{center}
{\includegraphics[height= 3.3111099996520in, width=5.696514566in]{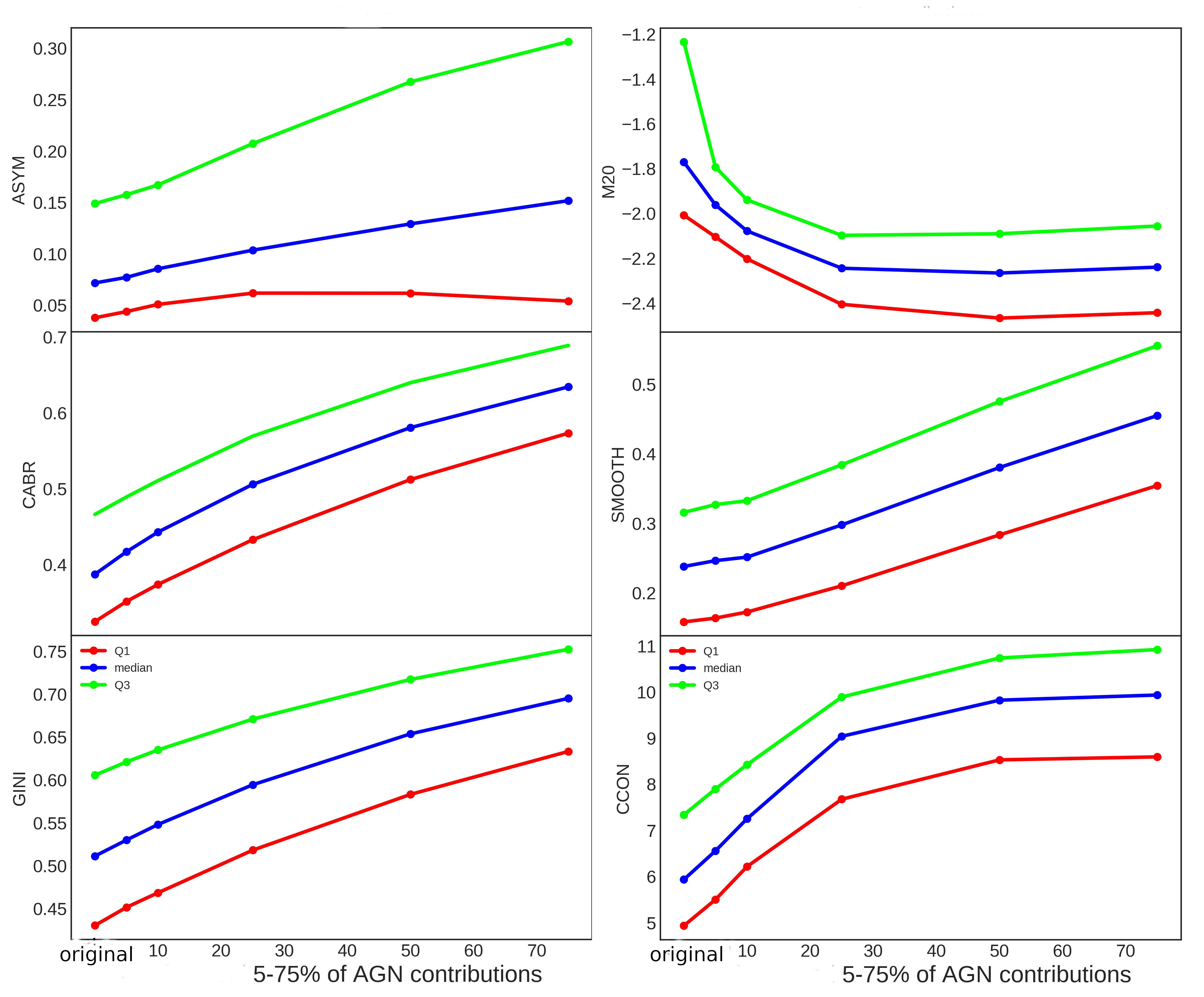}}
\caption{Effect of AGN contribution (5-75\%) on six morphological parameters at $z ∼\sim {0}$. The red line indicates quartile 1 (Q1), blue for median and lime
for quartile 3 (Q3)}.
\label{j}
\label{agnwhole2}
\end{center}
\end{figure}

\begin{figure}{}
\begin{center}
{\includegraphics[height= 3.6520in, width=5.6956522in]{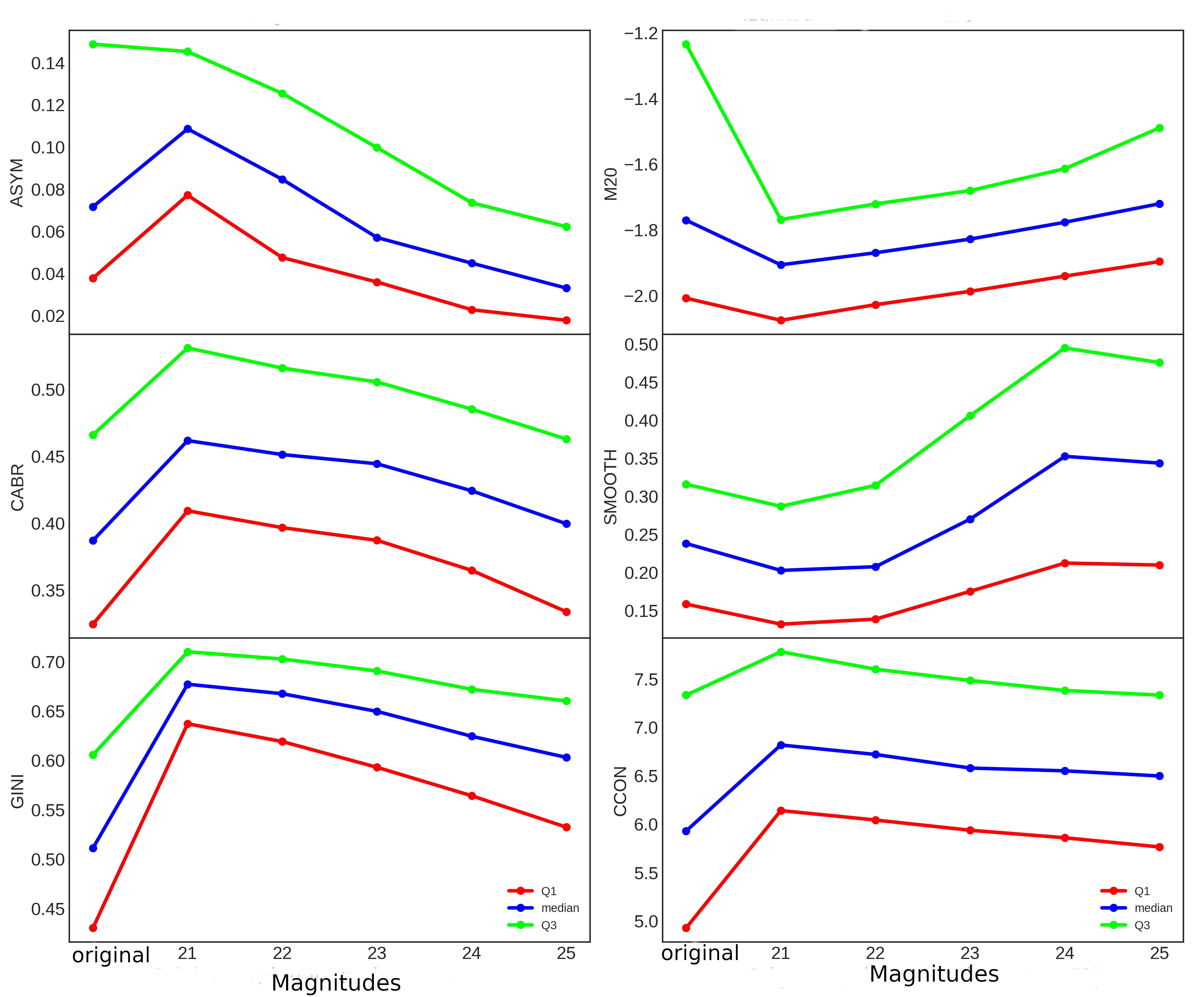}}
\caption{Comparison of six morphological parameters measured on the original sample (at $z$\,$\sim$\,0) and once moved to the redshift distribution in the COSMOS field and to fainter magnitudes (21 - 25).}
\label{j}
\label{whole1}
\end{center}
\end{figure}

\section{Conclusions}
The main objective of this work has been to study the effects of the AGN on the morphological
parameters of their host galaxies and to answer some of the fundamental questions still
open related to the effect of AGN on them. For the total sample, all parameters change significantly with the AGN contribution (with ASYM being less affect than others). At $z\,$=$\,0$ for most of the parameters up to a 10\% AGN contribution morphological classification will not be significantly affected, but for 25\% and higher late-type spirals follow the range of parameters typical of elliptical galaxies (\cite[Povi\'c et al. 2015]{Povic2015}).  When moving galaxies to higher $z$ and fainter mag, we conclude that the concentration indices CABR and CCON are less sensitive to redshift and magnitude than GINI, ASYM, M20, and SMOOTH. 

\section*{Acknowledgements}
TGW acknowledges the support from Bule Hora University. TGW, MP, and ZBA acknowledge financial support from the Ethiopian Space Science and Technology Institute (ESSTI) under the Ethiopian Ministry of Innovation and Technology (MoIT). TGW and JM acknowledge support by the grant CSIC I-COOP 2017,
COOPA20168. MP, JM, JP, and IMP acknowledge the support from the Spanish Ministry of
Science, Innovation and Universities (MICIU) through project AYA2016-76682C3-1-P and the State Agency for Research of the Spanish MCIU through the Center of Excellence Severo Ochoa” award to the Instituto de Astrof\'isica de Andaluc\'ia (SEV-2017-0709).

\end{document}